\documentclass[american]{article}
\usepackage[latin9]{inputenc}
\usepackage{textcomp}

\makeatletter
\newcommand{\lyxaddress}[1]{
\par {\raggedright #1
\vspace{1.4em}
\noindent\par}
}

\makeatother

\usepackage{babel}
\begin{document}

\title{Comment on \char`\"{}NO$\mathrm{_{x}}$ production in laboratory
discharges simulating blue jets and red sprites\char`\"{} by H. Peterson
et~al.}

\author{S. Nijdam$^{1}$, E.M. van Veldhuizen$^{1}$ and U. Ebert$^{1,2}$}

\maketitle

\lyxaddress{1) Department of Applied Physics, Eindhoven University of Technology,
Eindhoven, NL.}

\lyxaddress{2) Centrum Wiskunde \& Informatica, Amsterdam, NL.}

\subsection*{Introduction}

Knowing the importance of lightning for global atmospheric NOx production,
it is natural to ask how transient luminous events (TLEs) like sprites
and jets influence the NOx content of higher atmospheric layers. This
question has been addressed in the past in particular for sprites
as they are much more frequent than jets {[}\emph{Chen et al., }2008{]}.
Methods included calculation based estimates on the one hand {[}\emph{Sentman
et al,} 2008; \emph{Gordillo-Vazquez }2008; \emph{Neubert et al},
2008; \emph{Enell et al., }2008{]}, and observation based estimates
on the other hand {[}\emph{Arnone et al.} 2008{]}. \emph{Peterson
et al.} {[}2009{]} try to address the question through laboratory
experiments.

While such a study is certainly a desirable complement to calculations
and observations, it will never perfectly model atmospheric conditions
and therefore requires care when extrapolating to TLEs. Nevertheless,
we argue that \emph{Peterson} \emph{et al.} {[}2009{]} have made several
basic conceptual errors, and their study therefore cannot be used
for this purpose. The conceptual errors lie 1) in the assumption that
it would be sufficient to characterize a discharge only by pressure
and current, not distinguishing basically different discharge types,
2) in a wrong application of the similarity laws relating transient
cold discharges at different pressures, 3) in a confusion of individual
streamer channels within a sprite with carrot sprites as a whole and
4) in an overestimation of the (local) duration of sprite activity
by a few orders of magnitude. Furthermore, we have found a grave calculation
error regarding scaling, which has a direct influence on important
results reported in this paper.

\subsection*{Similarities between experiments and real TLEs}

One can use Townsend scaling to compare laboratory and real TLE discharges
when the two discharges are of a similar kind and both are dominated
by two-body collisions (as is the case for the propagating heads of
streamers and sprites). As was shown by \emph{Briels et al.} {[}2008{]},
it is possible to define properties like the reduced diameter (\emph{$n\times d$},
with \emph{n} the air density and \emph{d} the streamer diameter)
that are independent of pressure for a large pressure range.

Further extension of this scaling predicts how current, current density,
propagation velocity and more parameters of streamer-like discharges
scale with air density. All similarity laws are based on the mean
free path length of an electron between collisions with the neutral
gas molecules. Therefore, such scaling is only valid when the discharge
is dominated by such two-body collisions. This is the case for the
active tips of streamers and sprites, but not for sparks, leaders,
jets and lightning return strokes.

According to the similarity laws, if the same voltage is applied,
all length and time scales scale with the inverse neutral gas density
\emph{n}. Currents are independent of \emph{n} and therefore current
density scales with $n^{2}$. Velocities are independent of \emph{n}.\textbf{
}Townsend scaling for comparing streamer experiments with sprites
has been discussed e.g., by \emph{Pasko et al.} {[}1998{]}, \emph{Rocco
et al.} {[}2002{]}, \emph{Liu \& Pasko} {[}2004{]}, \emph{Ebert et
al.} {[}2006{]}, \emph{Luque et al.} {[}2007{]}, and \emph{Briels
et al.} {[}2008{]}. A recent review of the applications and limits
of Townsend scaling can be found in {[}\emph{Ebert et al.}, 2010{]}.

In contrast, \emph{Peterson} \emph{et al.} propose that their laboratory
experiments are similar to real TLEs because they have the same color,
pressure, current density and emission duration. In our view this
is incorrect for the following reasons:\\

i) It is well known that lightning at (nearly) atmospheric pressure
develops in several stages: coronas of cold streamer channels pave
the way of hot leader channels, and some leader channels later convert
into a very hot return stroke channel. All phases are clearly distinct
in their temperature, spectra and chemistry though they all evolve
at the same pressure. Therefore it is clearly insufficient to characterize
a discharge only by pressure as Peterson \emph{et al.} do. Up to now
it is frequently assumed that the return stroke channel would be the
main source of NOx production in lightning, but that (implicit) hypothesis
can be questioned. 

While it is commonly accepted that sprite discharges are a form of
streamer or corona discharges, \emph{Peterson} \emph{et al.} explicitly
mention that they want to avoid the occurrence of coronal discharges
in their experiments (their paragraph {[}19{]}). They claim that such
coronal discharges produce ozone and thus a higher proportion of NO$_{2}$.
But the laboratory equivalent of sprites are just corona (or streamer)
discharges.

\emph{Peterson} \emph{et al.} stress that their experimental discharges
are similar to real TLEs, but this is only similarity in a colloquial
sense and does not involve the similarity laws as discussed above.
They do not use scaling laws to compare densities, current densities
and dimensions of their discharges with TLEs. The alternative for
using scaling laws is to exactly replicate (a part of) a real TLE
discharge. In this case the experimental gas density should be equal
to that of the TLE discharge. In \emph{Petersons} sprite simulations
this is not the case.

From a plasma-technological point of view, it should be noted that
streamer coronas are used for more than a century to generate ozone
for various disinfection purposes\textbf{ }{[}\emph{van Veldhuizen},
2000{]} while glow discharges or sparks were clearly discarded for
this purpose. The purely empirical finding that streamer coronas produce
ozone in a very efficient manner is recently being supported by systematic
analysis and subsequent technological improvement. Through voltage
pulses lasting only several tens of nanoseconds, \emph{van Heesch
et al.} {[}2008{]} managed to convert more than 50\% of the electrical
input energy into O{*} radicals. The key to success was to power the
discharge only during the initial streamer phase while avoiding any
secondary streamers or gas heating. The underlying reason for this
efficiency is the electron energy distribution in the streamer head
with its high transient field; these electrons are very far from any
equilibrium as the reduced electric field $E/n$ is locally very high
{[}\emph{Morrow, }1985\emph{; Dhali and Williams,} 1987{]}. While
(hot and slow) spark and lightning-like discharges can produce about
$5\times10^{16}$~molecules joule$^{-1}$ of NO {[}\emph{Levine et
al., }1981{]}, (cold and fast) corona or streamer discharges like
the ones produced by \emph{Peterson et al.} are often more efficient
and can produce $7\times10^{17}$~molecules joule$^{-1}$ or more.\\

ii) \emph{Peterson} \emph{et al.} estimate the color of the discharges
from frames of a video of the discharge. In the images given in figure
7 of their paper, it is clear that the video is saturated in many
cases. In these cases, it is impossible to judge the color from the
video. Furthermore, just the fact that the color is similar is no
proof that the TLE discharge is similar.\\

iii) \emph{Peterson} \emph{et al.} claim that their current density
is similar to sprite current density, but do not proof this claim.
There is not much literature on sprite current density, but according
to \emph{Cummer} \emph{et al.}~{[}1998{]} the combined current of
all channels within a carrot sprite is 1.6--3.3~kA. The sprite-like
discharges by \emph{Peterson} \emph{et al.} have currents of order
10~A and a cross section of order $10^{-3}\,\mathrm{m}^{2}$. This
gives a current density of 10~kA/m$^{2}$. When we combine this current
density with the sprite current reported by \emph{Cummer} \emph{et
al.}, this would give a sprite diameter of less than 1~m. This is
lower than estimates of the diameter of one streamer channel in a
sprite discharge (which is of order 10--100~m). A real carrot sprite
consists of hundreds to ten-thousands of these streamers. Therefore,
we can conclude that the current density in a real sprite is probably
at least three orders of magnitude lower than in the laboratory discharges
by \emph{Peterson} \emph{et al. }Note that even if the current densities
would be similar, but the gas density is not, the discharges are not
similar in the sense of Townsend scaling as was discussed above.\\

iv) In sprite discharges, most light is emitted by the moving streamer
head. This occurs on short timescales (microseconds). The comparison
of the duration of their experiments by \emph{Peterson} \emph{et al.}
with the 1--2~ms and 0.53~ms duration of sprites from \emph{Pasko~}{[}2007{]}
is misleading. \emph{Pasko} mentions that the duration of order 1~ms
is a time integration over the motion of the sprite head (page S24
first paragraph). Locally, the sprite channel will only emit light
on a timescale of order 1~\textmu{}s (the length of the streamer
head is of order 10~m and its propagation velocity of order 10$^{7}$~m/s),
as was shown by fast imaging by \emph{McHarg et al. }{[}2007{]}.

The \textasciitilde{}1~ms discharge produced in the lab by \emph{Peterson}
\emph{et al.} has very different timescales, and therefore very different
chemistry. In this long duration discharge, multi-step reactions become
much more important, as there is enough time for reaction products
to react further. Such a discharge is close to equilibrium, while
a sprite (or streamer) is a transient discharge that is very far from
equilibrium. In a semi-equilibrium discharge, reactions of atoms,
ions, radicals and excited species become very important, while a
transient discharge is dominated by collisions of fast electrons with
neutral gas molecules in the ground state. The reaction products from
these collisions only start to play a significant role after the discharge
has passed and therefore they are never in equilibrium. Some examples
of the different timescales involved in the chemistry of a fast (nanosecond
to microsecond) pulse in air can be found in figure~10 from \emph{Eliasson
et al.} {[}1987{]} and figure~5 from \emph{van Veldhuizen et al.}
{[}1996{]}.

Furthermore, \emph{Peterson} \emph{et al.} use a damped oscillating
voltage and current to drive their discharge (see their Figure 5),
while real blue jets and sprites have a pulse-like current of a single
polarity. Proper comparison with sprites or blue jets is only possible
with a pulse forming network. Examples of such pulse forming networks
are C-supplies, transmission line transformers and Blumlein pulsers
{[}\emph{Briels} \emph{et al.,} 2006; \emph{Smith,} 2002{]}.\\

v) \emph{Peterson} \emph{et al.} mention in their paragraph {[}22{]}
that sprites consist of a series of streamers, each hundreds of meters
long and that this distance is required for the electrons to reach
equilibrium with the surrounding electric field. In reality, the streamers
in a sprite discharge are tens of kilometers long (and tens to hundreds
of meters wide). On the other hand, \emph{Li et al.} {[}2007{]} have
shown that the electron relaxation length at standard temperature
and pressure is about 1.5~\textmu{}m. By applying Townsend scaling,
we can see that at 65~km (0.1~mbar), the electron relaxation length
is about 1.5~cm and at 80~km (0.01~mbar) it is about 15~cm. Therefore
sprite lengths and electron relaxation lengths are not similar, but
differ by 4 to 5 orders of magnitude.\\

Summarizing, just the fact that there can be similarities between
carefully chosen laboratory experiments and TLEs does not mean that
all laboratory discharges represent a real TLE. There are many different
types of cold plasmas and they can have vastly different chemistries.
This can be determined by pressure, discharge duration, repetition
frequency, discharge current density and more, none of which the authors
prove to be equal between their discharges and real TLEs. The discharges
described by \emph{Peterson} \emph{et al.} do not represent real sprites
or blue jets. Especially the current duration and waveform in their
experiment are very different from those in a real TLE, even though
the pressure, color and gas composition may be similar to TLEs.

\subsection*{Comparison methods for NO$\mathrm{_{x}}$ production.}

\emph{Peterson} \emph{et al.} use two methods to compare the production
of NO$\mathrm{_{x}}$ by their laboratory discharges to real TLEs.
In their first method they compare the energy consumed by their discharges
with estimates of the energy of a real TLE. They take a lot of effort
to measure the energy dissipated by their laboratory discharge, but
then compare this to very rough estimates of TLE energies. These rough
estimates have been deduced from optical light emission measurements.

From recent observations by the ISUAL satellite mission, \emph{Kuo}
\emph{et~al.}~{[}2008{]} have calculated the average energy emitted
by a sprite discharge to be about 22~MJ. This is close to estimations
by \emph{Sentman et al. }{[}2003{]} which have a value of 1-10~MJ.
Both are much lower than the older rough estimates from \emph{Heavner}
\emph{et al.}~{[}2000{]} of 250~MJ -- 1~GJ that are used by \emph{Peterson
et al.}.

As discussed above, \emph{Van Heesch et al. }{[}2008{]} have shown
that O{*} radical production efficiency of a discharge can vary significantly
as function of discharge parameters.\emph{ Cooray} \emph{et al.} {[}2009{]}
show that also the NO$\mathrm{_{x}}$ production efficiency of electrical
discharges not only depends on the energy dissipated in the discharge
but also on the shape of the current waveform. This provides an explanation
for the different values of NO$\mathrm{_{x}}$ molecules/J obtained
by different researchers in different experiments. Thus, according
to \emph{Cooray} \emph{et al.}, energy dissipated in a discharge is
not suitable as the scaling quantity for extrapolating the laboratory
data to lightning flashes. We realize that \emph{Cooray et al.} {[}2009{]}
appeared later than the paper by \emph{Peterson et al.} but similar
reasoning was already presented in \emph{Cooray et al.} {[}2008{]}.

In our opinion, the statement by \emph{Cooray et al.} that one can
never extrapolate laboratory data to lightning flashes is too strong
and does not hold if the laboratory discharges are really similar
to the geophysical discharges. For discharges that obey Townsend scaling
{[}\emph{Ebert et al. }2010{]}, one can use scaling laws to extrapolate
laboratory data to geophysical discharges (e.g., comparing streamer
discharges with sprites), as long as enough knowledge about both discharges
is present.\\

The second comparison method by \emph{Peterson} \emph{et al. }uses
the geometric volume of the discharge to compare laboratory discharges
to real TLEs. Again, this method has problems with good data from
TLEs and depends a lot on a few field measurements.

In the case of sprites, \emph{Peterson} \emph{et al.} do not distinguish
clearly whether they use the complete volume of one carrot sprite,
the volume of one single sprite channel or the volume of all sprite
channels together. It seems that they have\emph{ }used the diameter
of a single sprite channel as estimated by \emph{Pasko} {[}2007{]},
but that they did not multiply this cross section by the number of
channels in a single sprite. They assume that \emph{Pasko}'s 'effective
diameter' takes this into account, while it does not, as it is the
diameter of a single streamer channel.\\

In conclusion, both comparison methods suffer from the same basic
problem: in contradiction to their claims, the laboratory experiments
by \emph{Peterson} \emph{et al.} are not similar to real TLEs and
in their application of scaling they make errors and leave uncertainties.

\subsection*{Calculation error.}

In table 3, \emph{Peterson} \emph{et al.} use the geometric method
to estimate the NO$\mathrm{_{x}}$ production by a blue jet, by assuming
that the production is proportional to volume. However, in their comparison
of the two geometries, they make a calculation error of 10$^{6}$.
The estimations of blue jet NO$\mathrm{_{x}}$ should be $1.7\times10^{28}$
to $6.4\times10^{29}$ instead of $1.7\times10^{22}$ to $6.4\times10^{23}$.
The value of $1.7\times10^{22}$ molecules production of NO$\mathrm{_{x}}$
per blue jet event is one of the most important results given in this
paper and it is quoted both in the abstract and in the conclusions.
Changing this value to its proper result of $1.7\times10^{28}$ would
change some quantitative results in the conclusions and abstract of
the paper by two to three orders of magnitude.

\subsection*{Conclusions}

Although we recognize that it is impossible to reproduce exact atmospheric
conditions in laboratory experiments, we think laboratory experiments
can definitely be used to simulate many aspects of TLEs. However,
this requires great care and a good understanding of the relation
between the laboratory conditions and the TLE. We argue that the arguments
by \emph{Peterson} \emph{et al.} regarding the similarities between
their experiments and real TLEs and their comparison methods for NO$\mathrm{_{x}}$
production are not supported by good proof or evidence and are in
many cases clearly wrong. They use a discharge voltage and current
that is a few orders of magnitude longer than sprite and blue jet
pulse durations and oscillates instead of having a fixed polarity.

The grave calculation error by a factor 10$^{6}$ regarding the geometric
estimation of NO$\mathrm{_{x}}$ production by a blue jet disqualifies
this paper further. This error affects an important value discussed
both in the abstract and in the conclusions of the paper.

For a proper laboratory measurement of NO$\mathrm{_{x}}$ production
from sprites, it would be worthwhile to study the efficiency of NO$\mathrm{_{x}}$
production in a similar way as \emph{van Heesch et al.} {[}2008{]}
have done for the O{*} production. In order to scale the measured
NO$\mathrm{_{x}}$ production to that of TLEs, one would need to combine
results of well chosen discharge experiments together with a review
of the density dependence of chemical models like used by \emph{Sentman
et al} {[}2008{]} and \emph{Gordillo-Vazquez }{[}2008{]}.

On the other hand, if one only wants to prove that jets and sprites
are no significant contributors to global NO$\mathrm{_{x}}$ production,
a simple calculation would suffice. The highest estimate of the energy
of one sprite or blue jet is about 1~GJ (which is probably an overestimate
as we argued above). If we assume that this 1~GJ is used to produce
NO with an enthalpy of formation of 90.29~kJ/mole at a 100\% conversion
efficiency (a clear overestimate), then each such TLE will produce
$6.7\times10^{27}$ molecules of NO. If we use NO$_{2}$ instead of
NO, this number would be about a factor three higher as its enthalpy
of formation is 33.1~kJ/mole.

In any case, the resulting maximum (overestimated) NO$\mathrm{_{x}}$
production is similar or slightly higher than the results of \emph{Peterson}
\emph{et al }and therefore would lead to the same conclusion that
TLEs do not significantly contribute to global NO$\mathrm{_{x}}$
production (when using the same assumptions of TLE occurrence and
global NO$\mathrm{_{x}}$ production). Scaling of results from laboratory
experiments to TLE energies can only give lower production estimates
which will not change this conclusion.

\section*{References}

 \providecommand{\natexlab}[1]{#1}   \providecommand{\doi}[1]{doi:\discretionary{}{}{}#1}
\begin{quote} Arnone, E., et~al. (2008), Seeking sprite-induced signatures in remotely sensed   middle atmosphere {NO}$_{2}$, \textit{Geophys. Res. Lett}, \textit{35},   L05,807, \doi{10.1029/2007GL031791}.
\end{quote}
\begin{quote} Briels, T. M.~P., J.~Kos, E.~M. van Veldhuizen, and U.~Ebert (2006), Circuit   dependence of the diameter of pulsed positive streamers in air, \textit{J.   Phys. D: Appl. Phys.}, \textit{39}, 5201, \doi{10.1088/0022-3727/39/24/016}.
\end{quote}
\begin{quote} Briels, T. M.~P., E.~M. van Veldhuizen, and U.~Ebert (2008), Positive streamers   in air and nitrogen of varying density: experiments on similarity laws.,   \textit{J. Phys. D: Appl. Phys.}, \textit{41}, 234,008,   \doi{10.1088/0022-3727/41/23/234008}.
\end{quote}
\begin{quote} Chen, A., et~al. (2008), Global distributions and occurrence rates of transient   luminous events, \textit{J. Geophys. Res. - Space Physics}, \textit{113}(A8),   A08,306, \doi{10.1029/2008JA013101}.
\end{quote}
\begin{quote} Cooray, V., M.~Becerra, and M.~Rahman (2008), On the {NO}$_{x}$ generation in   corona, streamer and low pressure electrical discharges, \textit{The Open   Atmospheric Science Journal}, \textit{2}, 176--180,   \doi{10.2174/1874282300802010176}.
\end{quote}
\begin{quote} Cooray, V., M.~Rahman, and V.~Rakov (2009), On the {NO}$_{x}$ production by   laboratory electrical discharges and lightning, \textit{Journal of   Atmospheric and Solar-Terrestrial Physics}, \textit{71}, 1877--1889,   \doi{10.1016/j.jastp.2009.07.009}.
\end{quote}
\begin{quote} Cummer, S.~A., U.~S. Inan, T.~F. Bell, and C.~P. Barrington-Leigh (1998), {ELF}   radiation produced by electrical currents in sprites, \textit{Geophys. Res.   Lett}, \textit{25}(8), 1281--1284, \doi{10.1029/98GL50937}.
\end{quote}
\begin{quote} Cummer, S.~A., J.~Li, F.~Han, G.~Lu, N.~Jaugey, W.~A. Lyons, and T.~E. Nelson   (2009), Quantification of the troposphere-to-ionosphere charge transfer in a   gigantic jet, \textit{Nature Geosci}, \textit{2}(9), 617--620,   \doi{10.1038/ngeo607}.
\end{quote}
\begin{quote} Dhali, S.~K., and P.~F. Williams (1987), Two-dimensional studies of streamers   in gases, \textit{J. Appl. Phys.}, \textit{62}(12), 4696--4707,   \doi{10.1063/1.339020}.
\end{quote}
\begin{quote} Ebert, U., C.~Montijn, T.~M.~P. Briels, W.~Hundsdorfer, B.~Meulenbroek,   A.~Rocco, and E.~M. van Veldhuizen (2006), The multiscale nature of   streamers, \textit{Plasma Sources Science and Technology}, \textit{15}, S118,   \doi{10.1088/0963-0252/15/2/S14}.
\end{quote}
\begin{quote} Ebert, U., S.~Nijdam, C.~Li, A.~Luque, T.~Briels, and E.~van Veldhuizen (2010),   Review of recent results on streamer discharges and discussion of their   relevance for sprites and lightning, \textit{J. Geophys. Res. - Space   Physics}, \textit{115}, A00E43, \doi{10.1029/2009JA014867}, in press.
\end{quote}
\begin{quote} Eliasson, B., M.~Hirth, and U.~Kogelschatz (1987), Ozone synthesis from oxygen   in dielectric barrier discharges, \textit{J. Phys. D: Appl. Phys.},   \textit{20}, 1421, \doi{10.1088/0022-3727/20/11/010}.
\end{quote}
\begin{quote} Enell, C.~F., et~al. (2008), {Parameterisation of the chemical effect of   sprites in the middle atmosphere}, \textit{Ann. Geophys.}, \textit{26},   13--27, \doi{10.5194/angeo-26-13-2008}.
\end{quote}
\begin{quote} Gordillo-Vazquez, F.~J. (2008), Air plasma kinetics under the influence of   sprites, \textit{J. Phys. D: Appl. Phys.}, \textit{41}(23), 234,016,   \doi{10.1088/0022-3727/41/23/234016}.
\end{quote}
\begin{quote} Heavner, M.~J., D.~D. Sentman, D.~R. Moudry, E.~M. Wescott, C.~L. Siefring,   J.~S. Morrill, and E.~J. Bucsela (2000), {Sprites, Blue Jets and Elves:   Optical Evidence of Energy Transport Across the Stratopause},   \textit{Geophysical Monograph--American Geophysical Union}, \textit{123},   69--82.
\end{quote}
\begin{quote} van Heesch, E. J.~M., G.~J.~J. Winands, and A.~J.~M. Pemen (2008), Evaluation   of pulsed streamer corona experiments to determine the {O}* radical yield,   \textit{J. Phys. D: Appl. Phys.}, \textit{41}(23), 234,015,   \doi{10.1088/0022-3727/41/23/234015}.
\end{quote}
\begin{quote} Kuo, C.~L., et~al. (2008), Radiative emission and energy deposition in   transient luminous events, \textit{J. Phys. D: Appl. Phys.}, \textit{41}(23),   234,014, \doi{10.1088/0022-3727/41/23/234014}.
\end{quote}
\begin{quote} Levine, J.~S., R.~S. Rogowski, G.~L. Gregory, W.~E. Howell, and J.~Fishman   (1981), Simultaneous measurements of {NO}$_{x}$, {NO}, and {O}$_{x}$   production in a laboratory discharge: {Atmospheric implications},   \textit{Geophys. Res. Lett.}, \textit{8}, 357--360,   \doi{10.1029/GL008i004p00357}.
\end{quote}
\begin{quote} Li, C., W.~J.~M. Brok, U.~Ebert, and J.~J. A.~M. van~der Mullen (2007),   Deviations from the local field approximation in negative streamer heads,   \textit{J. Appl. Phys.}, \textit{101}(12), 123305, \doi{10.1063/1.2748673}.
\end{quote}
\begin{quote} Liu, N., and V.~P. Pasko (2004), Effects of photoionization on propagation and   branching of positive and negative streamers in sprites, \textit{J. Geophys.   Res.}, \textit{109}, 1, \doi{10.1029/2003JA010064}.
\end{quote}
\begin{quote} Luque, A., U.~Ebert, C.~Montijn, and W.~Hundsdorfer (2007), Photoionization in   negative streamers: Fast computations and two propagation modes,   \textit{Appl. Phys. Lett.}, \textit{90}, 081,501.
\end{quote}
\begin{quote} McHarg, M.~G., H.~C. Stenbaek-Nielsen, and T.~Kammae (2007), Observations of   streamer formation in sprites, \textit{Geophys. Res. Lett.}, \textit{34},   L06,804, \doi{10.1029/2006GL027854}.
\end{quote}
\begin{quote} Morrow, R. (1985), Theory of negative corona in oxygen, \textit{Phys. Rev. A},   \textit{32}(3), 1799--1809, \doi{10.1103/PhysRevA.32.1799}.
\end{quote}
\begin{quote} Neubert, T., et~al. (2008), Recent results from studies of electric discharges   in the mesosphere, \textit{Surveys in Geophysics}, \textit{29}(2), 71--137,   \doi{10.1007/s10712-008-9043-1}.
\end{quote}
\begin{quote} Pasko, V.~P. (2007), Red sprite discharges in the atmosphere at high altitude:   the molecular physics and the similarity with laboratory discharges,   \textit{Plasma Sources Science and Technology}, \textit{16}, S13,   \doi{10.1088/0963-0252/16/1/S02}.
\end{quote}
\begin{quote} Pasko, V.~P., U.~S. Inan, and T.~F. Bell (1998), {Spatial structure of   sprites}, \textit{Geophys. Res. Lett}, \textit{25}, 2123--2126,   \doi{10.1029/98GL01242}.
\end{quote}
\begin{quote} Peterson, H., M.~Bailey, J.~Hallett, and W.~Beasley (2009), {NOx} production in   laboratory discharges simulating blue jets and red sprites, \textit{J.   Geophys. Res.}, \textit{114}, A00E07, \doi{10.1029/2009JA014489}.
\end{quote}
\begin{quote} Rocco, A., U.~Ebert, and W.~Hundsdorfer (2002), Branching of negative streamers   in free flight, \textit{Physical Review E}, \textit{66}, 35,102,   \doi{10.1103/PhysRevE.66.035102}.
\end{quote}
\begin{quote} Sentman, D.~D., H.~C. Stenbaek-Nielsen, M.~G. McHarg, and J.~S. Morrill (2008),   {Plasma chemistry of sprite streamers}, \textit{Journal of Geophysical   Research-Atmospheres}, \textit{113}(D11), D11,112,   \doi{10.1029/2007JD008941}.
\end{quote}
\begin{quote} Sentman, D.~D., et~al. (2003), Simultaneous observations of mesospheric gravity   waves and sprites generated by a midwestern thunderstorm, \textit{Journal of   Atmospheric and Solar-Terrestrial Physics}, \textit{65}(5), 537--550.
\end{quote}
\begin{quote} Smith, P.~W. (2002), \textit{Transient Electronics: Pulsed Circuit Technology},   Wiley, Chichester.
\end{quote}
\begin{quote} van Veldhuizen, E.~M. (2000), \textit{Electrical Discharges for Environmental   Purposes: Fundamentals and Applications}, Nova Science Publishers, New York.
\end{quote}
\begin{quote} van Veldhuizen, E.~M., W.~R. Rutgers, and V.~A. Bityurin (1996), Energy   efficiency of {NO} removal by pulsed corona discharges, \textit{Plasma Chem.   Plasma Process.}, \textit{16}, 227--247, \doi{10.1007/BF01570180}.
\end{quote}
\end{document}